# CAN OIL FLOAT COMPLETELY SUBMERGED IN WATER?

Saurabh Nath
Mechanical Engineering Department
Jadavpur University
Kolkata, West Bengal, India
saurabh0692@gmail.com

Anish Mukherjee
Electrical Engineering Department
Jadavpur University
Kolkata, West Bengal, India
anishmukherjee47@gmail.com

Souvick Chatterjee
Mechanical Engineering Department
Jadavpur University
Kolkata, West Bengal, India
souvickchat@gmail.com

## ABSTRACT

Droplet formation in a system of two or more immiscible fluids is a celebrated topic of research in the fluid mechanics community. In this work, we propose an innovative phenomenon where oil when injected drop-wise into a pool of water moves towards the air-water interface where it floats in a fully submerged condition. The configuration, however, is not stable and a slight perturbation to the system causes the droplet to burst and float in partially submerged condition. The droplet contour is analyzed using edge detection. Temporal variation of a characteristic length of the droplet is analyzed using MATLAB image processing. The constraint of small Bond Number established the assumption of lubrication regime in the thin gap. A brief theoretical formulation also showed the temporal variation of the gap thickness.

## INTRODUCTION

The motion of a fluid droplet in an immiscible fluid medium towards the fluid-fluid interface is an interesting phenomenon both from fundamental and application point of view. While such a problem is classically challenging, it finds application in fields of environment and microfluidics. The interface gets deformed and the motion of the droplet is significantly altered. An interesting effect of this interaction is the fully submerged floatation of a lighter fluid in a denser one when the latter is released drop-wise into the pool of former through a horizontal nozzle. The study can be also exemplified as a simple model of drop coalescence. Drop coalescence can be both a favorable or non-favorable condition depending on the physical scenario (Mohamed-Kassim and Longmire, 2004). Several studies have addressed the coalescence phenomenon using experimental (Ban et al., 2000, Paulsen et al., 2011) as well as numerical (Ban et al., 2000, Paulsen et al., 2011) and analytical (Blanchette et al., 2009) techniques. Thus by studying the parameters affecting the drop coalescence with its bulk phase, we can effectively determine the applicability of such a system where a drop of oil moves towards an interface of water and a thin oil film.

Bond Number, defined as the ratio of buoyant forces to the surface tension, is an important parameter in problems of this kind. This parameter plays a crucial role in the stability of the droplet (Bezdenejnykh et al., 1992). In most cases, the problem is treated as a film drainage problem with little attention on the shape of the fluid drop. Also effect of fluid properties like viscosity and kinematic properties in the form of initial velocities have been incorporated in few of the studies. Such a detailed work had been done by Chi and Leal (1989) by taking into account viscous effects on the shape of the drop as well as the initial shape of the film at the interface. Effect of Bond Number has also been dealt analytically by Manga and Stone (1995) for low Reynolds Number motion of bubbles.

The instability of the droplet is because of film drainage. The entrapped water in the gap drains slowly leading to spreading of the oil drop. A very detailed analytical description on the film drainage problem and its effect on droplet coalescence has been provided by Jones and Wilson (1978). Yiantsios and Davis (1990) took into account small Bond Number and used lubrication theory to explain the drainage of the thin drop between the drop and the interface. Also, noteworthy, both cases of solid and deformable interface have been considered in this work. The draining has been further studied by P.D. Howell (1999) where the drainage has been observed as the responsible phenomenon for rupture of the film ,i.e. instability of the droplet. They used lubrication analysis to explain draining of a bubble at the interface of molten gas and air. An interesting experimental analysis on film drainage of a deformable oil droplet using ellipsometry/reflectometry has been shown by Cubaud et al. (2001). The viscous stresses for such draining films have been estimated by Vaughn and Slattery (1995) for both mobile and immobile interfaces. They used a technique postulated by Bird et al. (1977) and compared the same with lubrication theory.

The liquid-liquid interface differs from the solid-liquid one in boundary conditions which leads to interesting phenomena. The shape of a deformable interface is reported to be specified by the normal component of a momentum balance jump (Slattery et al., 2006). The general lubrication theory resembling squeezing flow between two parallel disks fails in certain aspects and is addressed in this work. Shopov and Minev (1992) studied breakage of interface due to the buoyancy-driven motion of a deformable drop. A numerical model has been proposed by Gellar et al. (1986) where a rigid spherical body moves towards a deformable fluid-fluid interface. Effects of various parameters in the form of viscosity ratio, density difference and interfacial tension between two fluids have been addressed in this work. But the phenomenon of fully submerged oil droplets in a pool of water is an interesting observation and requires good investigation. In this work, under room conditions the average time period for which such a droplet floats in fully submerged condition is calculated. Analysis of the droplet profile with a characteristic length as a quantitative measure led to the determination of temporal variation of the droplet shape.

Such a phenomenon may possibly be encountered in chemical engineering processes like liquid-liquid extraction, where a liquid is dispersed in an immiscible liquid medium in the form of droplets. The large interfacial area thus created helps the efficient passage of the solute from one liquid to the other. The liquids need to be separated again, at the end of the process. Herein contamination occurs in each liquid by small droplets of the other. Such contamination needs be prevented. Two bulk layers are formed by the two liquids, the lighter liquid being uppermost. However, each contains small drops of the other approaching the interface under the influence of buoyancy forces. The speed and efficiency with which these drops can be removed is often a key factor in the overall efficiency of the process.

**EXEEPRIMENTAL SET-UP**

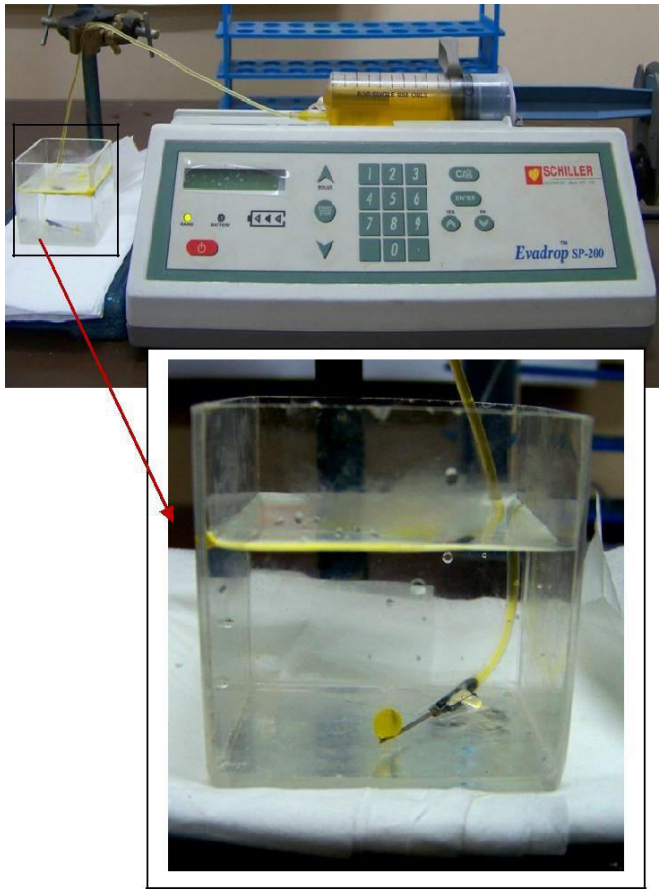

Fig. 1: Experimental Setup

The experimental setup with all the equipment is shown in Fig. 1. As can be seen, a 50 ml Dispovan Syringe is filled with mustard oil and with the aid of a syringe infusion pump, the oil is pumped into a pipe. A 24G Butterfly Needle is attached downstream of the pipe which drop-wise releases the oil into an Acrylic Sheet container filled with distilled water. The mouth of the butterfly needle is fixed to the bottom of the container using double sided tapes. The use of the syringe pump ensures a steady flow rate through the needle throughout the course of the experiment, thereby resulting in the formation of oil bubbles of equal volume at the mouth of the butterfly needle. The syringe pump is kept in a vertical position to provide an air cushion so that the small jerks caused by the progress of the piston have no effect on the dynamics of the bubble before detachment and it may be reasonably assumed that the bubble starts from rest following detachment. A 24G butterfly needle is used since it has the least aperture radius amongst the available needles, and thus gives the smallest volume of the bubble.

A flow-rate of 10ml/hr is set in the syringe pump. This small flow-rate ensures that the disturbance caused by the oil flow in the volume and dynamics of the oil bubble is reasonably small and sufficient time is given between the formations of subsequent bubbles. The first few instances are neglected due to the invariable presence of trapped air bubbles within the oil bubble. The phenomenon is recorded using a 12 Megapixel Digital Single Reflex (DSLR) Camera with a 55-300 lens at a standard definition of 640 x 416 pixels. The captured video is then processed to separate the frames which are then individually processed to get the edges of the bubbles, thereby facilitating a better understanding and clearer view of the dynamics of the bubble.

**RESULTS**

The mobile interface allows deformation of the same when the oil droplet strikes it.

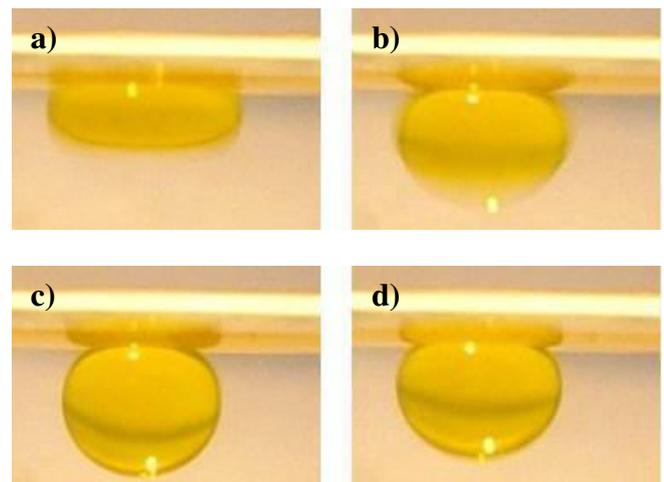

Fig. 2: *(Frame a-d)* Images of Oil-Droplet striking the mobile fluid-fluid interface

A lighter fluid (mustard oil) when injected drop-wise in a pool of heavier fluid (water), the oil droplet rises and strikes the oil-water mobile interface. As can be seen in Fig. 2, the droplet bounces a little on striking the mobile interface and then stabilizes. However, the stabilization holds for a small amount of time, after which disintegration of the droplet owing to film drainage is observed. As shown by Kocarkova et al. (2010), the Bond Number, defined as,

$$Bo = \frac{\Delta \rho \, g \, D^2}{\sigma} \quad (1)$$

plays an important role in the stability of such a droplet. The '*D*' in the above equation is a characteristic length of the problem and the maximum width of the droplet is considered for this purpose. Edge detection algorithm is used in a JAVA based image processing software called **ImageJ**, to obtain the outer profile of the droplet in a particular configuration.

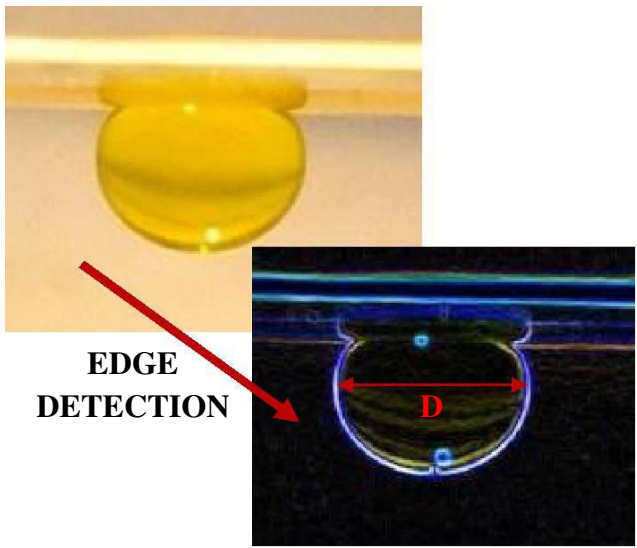

**EDGE DETECTION**

Fig. 3: Edge Detection Algorithm producing outer contour of an oil droplet

Such an image with highlighted edges proves helpful for quantitative analysis. The entire video of buoyant oil droplet striking the interface stabilizing for a while and subsequent draining of the oil is broken into frames. Edge detection algorithm is applied on all such frames to obtain the sharp outer contour profile.

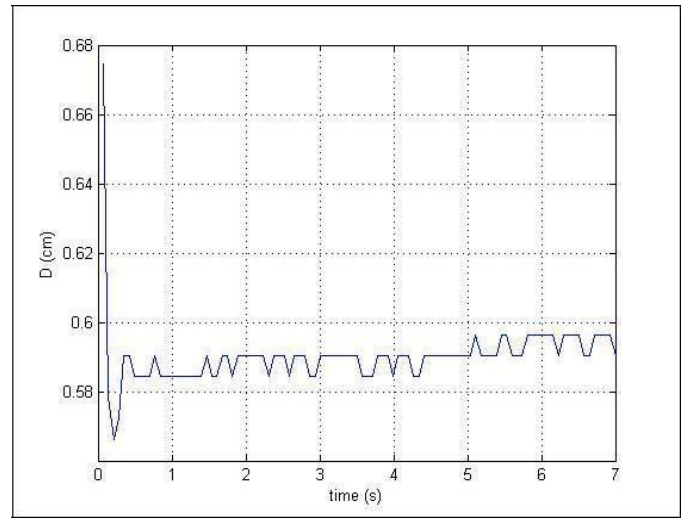

Fig. 4: Variation of the characteristic length of droplet with time

An image processing technique is used in MATLAB where D, the characteristic length scale, defined as the maximum width of the droplet is calculated for all such images. A scale attached to the setup provided the calibration parameters such that 1 cm corresponds to 166 pixels. The initial peak value of D corresponds to the first strike of the droplet at the interface (Fig. 2a), after which the characteristic length stabilizes at a steady value of 0.58-0.6 cm.

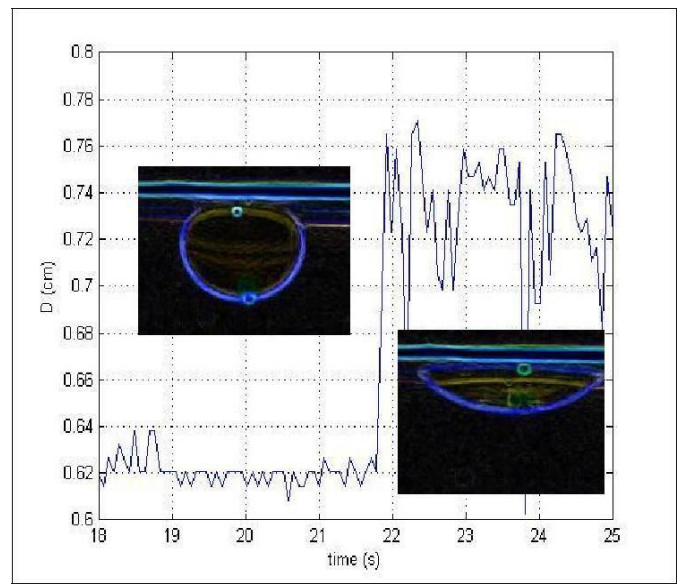

Fig. 5: Sudden change of characteristic length with decrease in stability

Figure 5 shows the effect of film drainage on the characteristic length. There is a sudden rise in the magnitude of D because of the shape change which also signifies the sudden collapse of the droplet. Also noteworthy in this figure, the instability in the system after the profile change is apparent from the variation in the magnitude of characteristic length. Thus a quantitative estimate of the instability can be obtained from this plot. Time is plotted along x-axis which suggests the droplet to remain in a stable configuration for about 22 seconds before the sudden collapse. The error in the experimental results is calculated by repeating the experiment and the error magnitude is found to remain within 3%.

An important parameter in such problems is the capillary length defined as:

$$l_{cap} = \sqrt{\frac{\sigma}{\Delta \rho g}} \qquad (2)$$

It is well known that the stability of a droplet/bubble is dependent on Bond Number (*Bo*), specifically the relationship between the cahracteristic length and capillary length ($l_{cap}$) such that:

$l_{characteristic}$< $l_{cap}$ => stable spherical shape of the bubble/droplet

$l_{characteristic}$ > $l_{cap}$ => deformation of shape of the bubble/ droplet

The parameters in this work led to a smaller *Bo* compared to $l_{cap}$ and hence the final film drainage and collapsing of the droplet validates the above inequality.

**"Eclipse" of a Droplet**

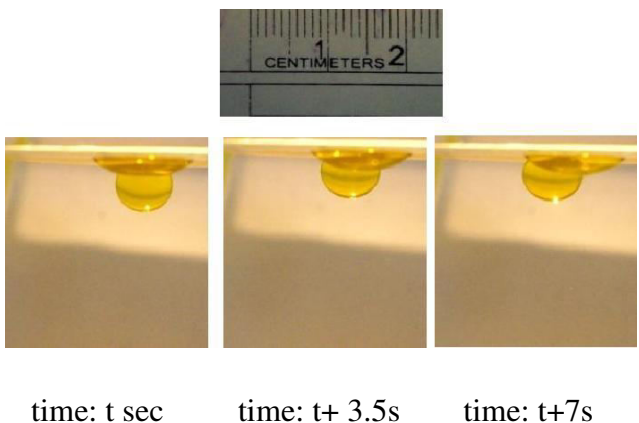

time: t sec        time: t+ 3.5s       time: t+7s

Fig. 6: Temporal Variation of a droplet
eclipse *(top)* scale bar

An interesting phenomenon observed while studying such submerged floating oil droplet in water is the eclipse of a droplet. This occurs when a fresh oil droplet strikes the interface at the position of a partially collapsed droplet. As can be seen in Fig. 6, the new droplet faces a "partial eclipse" such that view of part of it obstructed by the older droplet. The noteworthy thing in this context is the inability of altering the stability of either of the droplet by this phenomenon. The gradual temporal movement of the new spherical droplet provides small perturbation to the interface which fails to accelerate the instability of the droplets.

**ANALYTICAL FORMULATION**

The instability of the droplet causing the sudden partial collapse followed by complete disintegration is a result of film drainage, the hydrodynamics of which is an interesting topic of research. In this work, a brief analytical formulation is presented using lubrication theory assumption on continuity and momentum equation to estimate the temporal variation of the film thickness between the droplet and the mobile interface.

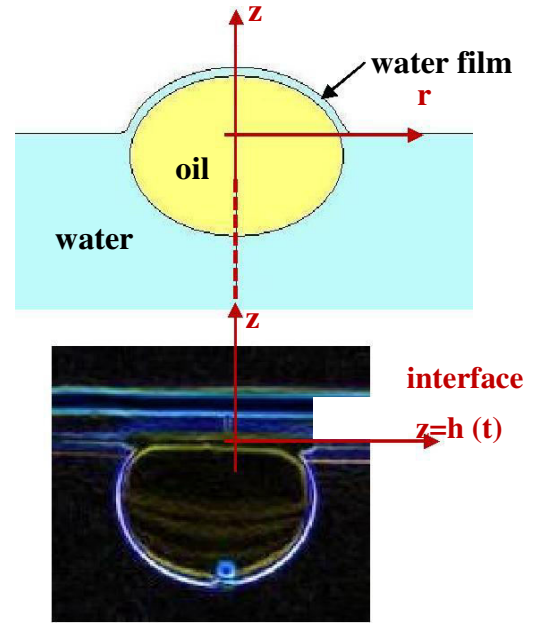

Fig. 7: The Axes System *(top)* schematic superimposed on a submerged oil droplet

Figure 7 shows a thin liquid film is entrapped between the fluid-fluid mobile interface and the oil droplet. Based on this, a lubrication theory assumption can be made to such a problem as has been shown by Srinivas et. al.(1999) and is adapted in this formulation.

The continuity equation for such a system can be written as:

$$\frac{1}{r}\frac{\partial}{\partial r}(rv_r)+\frac{\partial v_z}{\partial z}=0 \qquad (3)$$

where $v_r = v_r(r,z,t)$ and $v_z = v_z(z,t)$ as mentioned by Bird et al. (1977).

The lubrication theory assumption of z-direction velocity gradient much larger than radial velocity gradient leads to the simplified momentum equations:

$$-\frac{\partial p}{\partial r}+\mu\frac{\partial^2 v_r}{\partial z^2}=0 \qquad (4)$$

$$-\frac{\partial p}{\partial z}+\mu\frac{\partial^2 v_z}{\partial z^2}=0 \qquad (5)$$

The mobile fluid-fluid interface leads to the boundary conditions in the form of:

At $z = 0$ (i.e r-axis) $\quad \frac{\partial v_r}{\partial z}=0$ and $v_z = 0$ (6)

At $z = h(t)$ $\quad \frac{\partial v_r}{\partial z}=0$ and $v_z = \frac{dh(t)}{dt}$ (7)

Solving the governing equations (Eq. 3-5) with the boundary conditions (Eq. 6-7) leads to the velocity components:

$$v_r = -\frac{1}{2}\frac{r}{h}\frac{dh}{dt}\text{ and } v_z = \frac{z}{h}\frac{dh}{dt}$$

As shown by Vaughn and Slattery (1995), this velocity components lead to an expression for the normal force on the upper interface given by:

$$F_z = -\Delta p\pi R^2 - 2\pi R^2\mu\frac{1}{h}\frac{dh}{dt} \qquad (8)$$

where $\Delta p$ = difference in hydrostatic pressure and pressure inside the bubble and $R = \frac{D}{2}$, i.e. half of the characteristic length defined earlier. Using an order of magnitude method, in this work, the variation of the film thickness (h) with time has been analyzed. Hence, the scale of the parameters in Eq. 8 needs to be estimated. For such estimation, assuming a spherical shape of the droplet it can be written: $\Delta p \sim \frac{\sigma}{R}$ .

Also, Equation 8 can be re-written as:

$F_z = -C_1 - \frac{C_2}{h}\frac{dh}{dt}$ where $C_1 = \Delta p\pi R^2$ and $C_2 = 2\pi R^2\mu$

Hence, $C_2\int_{h_0}^{h}\frac{dh}{h} = -(F_z+C_1)\int_{0}^{t}dt$

$\Rightarrow C_2\ln(\frac{h}{h_0}) = -(F_z+C_1)t$

$\Rightarrow h = h_0 e^{-at}$ (9) Where $a = \frac{F_z+C_1}{C_2}$

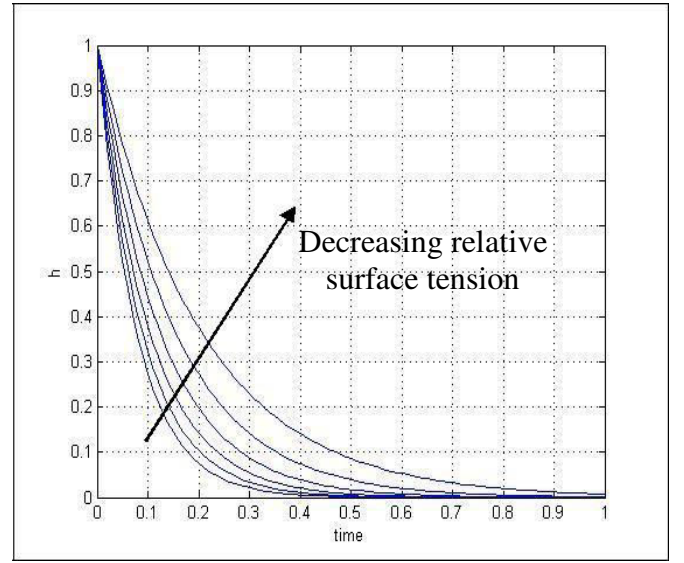

Fig. 8: Variation of film thickness for varying surface tension

Therefore, for a fully mobile interface, the film thickness decays exponentially with time. Now,

$C_1 = \Delta p \pi R^2 \sim \pi \sigma R$. Also as initially $\frac{dh}{dt} = 0$, $F_z = C_1$. This along with the value of $C_2$ provides an estimate of "a" in Eq. 9.

Such an order of magnitude analysis allows us to plot the variation of the film thickness with time. Using the more consistent attainable diameter of 0.59 cm as the characteristic length scale, the exponential decay nature of the film thickness is shown in Fig. 8. Also shown in the figure is the effect of surface tension on the variation of this parameter. As the relative surface tension of the two fluids involved in the phenomenon is decreased, the rate of decay of the film thickness decreases. Hence, the enhanced ability of surface tension in increasing the instability of the buoyant droplet is established.

## OBSERVATIONS

In a previous work by Sreenivas and De (1999) they have shown that water droplets can float on water when there is an air film entrapped. The phenomenon that has been observed here is in essence the same with the lighter liquid (oil) floating completely submerged in water due to the entrapment of a water film between air and the oil drop itself. This shows that film drainage may cause an 'apparent' floatation in such non-intuitive configurations in many cases where it is not expected at all. As is evident from the analytical formulation the phenomenon is dominated by surface tension.

## CONCLUSIONS

An innovative phenomenon related to droplet dynamics is observed and analyzed in this work. Studies involving oil-water combination comprise a very intriguing and fundamental topic of research but most of the studies have focused on a system with oil floating on water. Hence, a very novel approach is used in this study when oil is injected drop-wise into a pool of water from beneath the oil-water interface. This led to the formation of fully submerged oil droplets at the fluid-fluid mobile interface. The temporal variation of these droplets is studied herewith.

Using edge detection algorithm the profile of such a droplet is analyzed.

The maximum width of a droplet is identified as the characteristic length, temporal variation of which is monitored using image processing in MATLAB. The initial steady nature of this parameter is found to change to a sudden oscillating pattern signaling the onset of instability. Also, the magnitude of the characteristic length increases suddenly suggesting partial collapse of the droplet. Also, sudden interesting things like partial eclipse of a droplet is observed and reported in this work. This happens when a fresh droplet strikes the interface at the position where a partially collapsed earlier droplet is present. Such close interaction between two droplets fails to decrease the stability of either of the droplets.

Also, in this work, a brief analytical formulation is presented to show the temporal variation of film thickness. Lubrication theory assumption along with some scale analysis has been used to analytically study this kind of a film drainage problem. The established nature of exponential decay of the film thickness is shown along with the effect of varying relative surface tension between the two fluids. This kind of study is useful both as fundamental research and applicability in certain areas involving multicomponent fluid flow. Hence, the present work showing an interesting phenomenon can prove useful for future detailed study of droplet dynamics in similar fluid-fluid systems.

## ACKNOWLEDGEMENTS

Experiments were carried out with the micro-syringe pump in the Optical Diagnostics Laboratory, Mechnaical Engineering Department, Jadavpur University. The authors are indebted to Prof. Swarnendu Sen and Prof. Ranjan Ganguly.

## NOMENCLATURE

| | |
|---|---|
| Bo | Bond Number |
| $\rho$ | Density |
| g | Acceleration due to gravity |
| D | Characteristic Length of droplet |
| $\sigma$ | Surface Tension |
| $l_{cap}$ | Capillary Length |
| $v_r$ | Radial Velocity (in r direction) |

$v_z$     Axial Velocity (in z direction)
h       film thickness
p       Pressure
$\mu$     Dynamic Viscosity
$h_0$     Initial Film Thickness

# REFERENCES

Ban T, Kawaizumi F, Nii S, Takahashi K. 2000. Study of drop coalescence behavior for liquid–liquid extraction operation. Chemical Engineering Science (55): 5385-5391.

Bezdenejnykh NA, Meseguer J, Perales JM. 1992. Experimental analysis of stability limits of capillary liquid bridges. Physics of Fluids A: Fluid Dynamics (4): 677-680.

Bird RB. 1977.Dynamics of polymeric liquids: Wiley.

Blanchette F, Messio L, Bush JWM. 2009. The influence of surface tension gradients on drop coalescence. Physics of Fluids (21): 072107-072110.

Chi BK, Leal LG. 1989. A theoretical study of the motion of a viscous drop toward a fluid interface at low Reynolds number. Journal of Fluid Mechanics (201): 123-146.

Cubaud T, Gee M, L., Fermigier M, Goodall D, G., Jenffer P, Stevens G, W., 2001. Imagerie de profils de gouttes lors du drainage d'un film entre une goutte et une interface. Oil & Gas Science and Technology - Rev. IFP (56): 33-40.

Geller AS, Lee SH, Leal LG. 1986. The creeping motion of a spherical particle normal to a deformable interface. Journal of Fluid Mechanics (169): 27-69.

Howell PD. 1999. The Draining of a two-dimensional bubble. Journal of Engineering Mathematics (35): 251-272.

Jones AF, Wilson SDR. 1978. The film drainage problem in droplet coalescence. Journal of Fluid Mechanics (87): 263-288.

Koˇcárková H, Pigeonneau F, Rouyer F. 2010.Film drainage between bubble and fluid interface Particle Bubble and Drop Dynamics. 7th International Conference on Multiphase Flow - ICMF 2010 Proceedings. University of Florida: International Conference on Multiphase Flow (ICMF).

Manga M, Stone HA. 1995. Low Reynolds number motion of bubbles, drops and rigid spheres through fluid–fluid interfaces. Journal of Fluid Mechanics (287): 279-298.

Mohamed-Kassim Z, Longmire EK. 2004. Drop coalescence through a liquid/liquid interface. Physics of Fluids (16): 2170-2181.

Paulsen JD, Burton JC, Nagel SR. 2011. Viscous to Inertial Crossover in Liquid Drop Coalescence. Physical Review Letters (106): 114501.

Shopov PJ, Minev PD. 1992. The unsteady motion of a bubble or drop towards a liquid-liquid interface. Journal of Fluid Mechanics (235): 123-141.

Slattery JC, Sagis L, Oh ES. 2006.Interfacial Transport Phenomena: Springer.

SREENIVAS KR, DE PK, ARAKERI JH. 1999. Levitation of a drop over a film flow. Journal of Fluid Mechanics (380): 297-307.

Vaughn MW, Slattery JC. 1995. Effects of Viscous Normal Stresses in Thin Draining Films. Industrial & Engineering Chemistry Research (34): 3185-3186.

Yiantsios SG, Davis RH. 1990. On the buoyancy-driven motion of a drop towards a rigid surface or a deformable interface. Journal of Fluid Mechanics (217): 547-573.